\documentclass[a4paper,11pt]{article}
\pdfoutput=1
\usepackage{jheppub}

\usepackage{graphicx}
\usepackage{placeins}
\usepackage{url}
\usepackage{multirow}
\usepackage{paralist}
\usepackage{amssymb}
\usepackage{amsmath}
\usepackage{xspace}
\usepackage{booktabs}
\usepackage[symbol]{footmisc}

\usepackage[caption=false]{subfig}

\usepackage{color}
\definecolor{darkred}{rgb}{1.0,0.1,0.1}
\definecolor{darkgreen}{rgb}{0.1,0.7,0.1}
\definecolor{darkblue}{rgb}{0.1,0.1,1.0}

\begin{document}

\title{Unbinned Inference with Correlated Events}

\affiliation[a]{Physics Division, Lawrence Berkeley National Laboratory, Berkeley, CA 94720, USA}
\affiliation[b]{Department of Physics, University of California, Berkeley, CA 94720, USA}
\affiliation[c]{Department of Physics and Astronomy, University of California, Riverside, CA 92521, USA}
\affiliation[d]{Berkeley Institute for Data Science, University of California, Berkeley, CA 94720, USA}

\author[a, b]{Krish Desai,} \emailAdd{krish.desai@berkeley.edu}

\author[c]{Owen Long,}
\emailAdd{owenl@ucr.edu}
\author[a,d]{and Benjamin Nachman}
\emailAdd{bpnachman@lbl.gov}

\abstract{
Modern machine learning has enabled parameter inference from event-level data without the need to first summarize all events with a histogram.  All of these unbinned inference methods make use of the fact that the events are statistically independent so that the log likelihood is a sum over events.  However, this assumption is not valid for unbinned inference on unfolded data, where the deconvolution process induces a correlation between events.  We explore the impact of event correlations on downstream inference tasks in the context of the OmniFold unbinned unfolding method.  We find that uncertainties may be significantly underestimated when event correlations are excluded from uncertainty quantification.
}

\maketitle


\section{Introduction}
\label{sec:intro}

Accurate and precise parameter estimation is a central task in high-energy physics data analysis.  Traditional approaches usually involve summarizing high-dimensional, event-level data into a small number of observables, whose differential cross sections are approximated with histograms.  Best fit parameters and confidence intervals are then extracted using first-principles-based or Monte Carlo simulation-based templates using binned likelihood approaches.   While effective, these methods can suffer from significant information loss and potential biases due to the dimensionality reduction and the binning.  Furthermore, classical methods cannot be employed to study the multidifferential cross section of many observables simultaneously due to the increasing difficulty of binning effectively as the number of dimensions increases.  Even if a simultaneous binned measurement was performed, it is not feasible to extract the differential cross section of lower-dimensional observables computed from the higher-dimensional phase space.

The advent of modern machine learning techniques has enabled unbinned inference methods that operate directly on event-level data without the need for dimensionality reduction or histogramming~\cite{Cranmer_2020}.  These methods exploit the full information content of the data, leading to improved sensitivity and reduced biases in parameter estimation. A key assumption underlying these unbinned methods is the statistical independence of events, which allows the joint likelihood to be expressed as a product over individual event likelihoods, or equivalently, the log-likelihood as a sum over events:

\begin{equation}
\label{eq:unbinned}
\ln \mathcal{L}(\theta) = \sum_{i=1}^{N} \ln p(x_i |\theta),
\end{equation}
where  $x_i$  denotes the observed data for event  $i ,  \theta$  represents the parameters of interest, and  $p(x_i \,|\, \theta)$  is the probability density function for event  $i$ given  $\theta$.  The widely-used asymptotic formulae~\cite{Cowan:2010js} for likelihood-based statistics apply also to the unbinned case and can significantly accelerate the calculation of confidence intervals and $p$-values.

Such methods have been effectively applied to detector-level data~\cite{ATLAS:2024jry}, where the assumption of the statistical independence of events holds. However, this assumption of event independence does not hold in the context of unfolded data.  Unfolding is the process of statistically removing detector distortions to produce cross-section measurements at the particle-level.  These results are independent of the detector and can thus be analyzed outside of experimental collaborations.  Unfolding is an ill-posed inverse problem that requires regularization techniques~\cite{Cowan:2002in,Blobel:2203257,doi:10.1002/9783527653416.ch6,Brenner:2019lmf} and introduces correlations between events due to the deconvolution process. These correlations violate the independence assumption.  Nonetheless, unfolded data have been used extensively to measure fundamental parameters, fit parton distribution functions~\cite{Ethier:2020way}, tune parton shower Monte Carlo simulators~\cite{Buckley:2009bj}, and search for physics beyond the Standard Model~\cite{Butterworth:2016sqg}.  All of these studies use binned data, where correlations can be properly accounted with a covariance matrix.  The usual Gaussian limit of Eq.~\ref{eq:unbinned} results in a $\chi^2$ test statistic.  With bin-by-bin correlations, the statistic most often used is
\begin{equation}
\label{eq:chi2}
\chi^2_{\text{full}}(\theta) = (D - P(\theta))^\top \Sigma^{-1} (D - P(\theta)),
\end{equation}
where  $D$  is the vector of unfolded data counts,  $P(\theta)$  is the vector of predicted counts from the model at parameters  $\theta$, and  $\Sigma$  is the covariance matrix encoding statistical and systematic uncertainties.  As with Eq.~\ref{eq:unbinned}, asymptotic formulae apply to Eq.~\ref{eq:chi2} for deriving confidence intervals and $p$-values.

Recent innovations from machine learning have enabled unbinned unfolding~\cite{Arratia:2021otl,Butter:2022rso2,Huetsch:2024quz}.  These methods are based on discriminative~\cite{Andreassen:2019cjw,Andreassen:2021zzk,Pan:2024rfh} or generative methods~\cite{Datta:2018mwd,Howard:2021pos,Diefenbacher:2023wec,Butter:2023ira,Bellagente:2019uyp,Bellagente:2020piv,Vandegar:2020yvw,Backes:2022sph,Leigh:2022lpn,Ackerschott:2023nax,Shmakov:2023kjj,Shmakov:2024gkd} and there are now also experimental results with some of these approaches~\cite{H1:2021wkz,H1prelim-22-031,H1:2023fzk,H1:2024mox,LHCb:2022rky,ATLAS:2024xxl,ATLAS:2025qtv,CMS-PAS-SMP-23-008,Song:2023sxb,Pani:2024mgy}.  One can bin the unbinned results and use Eq.~\ref{eq:chi2} for parameter estimation, but we are not aware of an analog to Eq.~\ref{eq:unbinned} that accounts for correlations.  The goal of this paper is explore how unbinned unfolded results can be used for parameter estimation.  While there are a growing number of unbinned experimental results, the methods and publications are sufficiently new that there are currently no parameter extractions using these data.  Our study will establish some best practices for using unbinned unfolded data for parameter estimation as well as new research questions for improving these protocols in the future.

In particular, we are interested in a number of questions: (1) is there an advantage in precision for binning after the unfolding instead of before? (2) is there any effect on the inference precision if correlations are ignored when creating the test statistic? and (3) if we ignore correlations and use Eq.~\ref{eq:unbinned} for the unbinned data, are the asymptotic uncertainties still valid?   For (1), we will compare classical binned unfolding with an unbinned approach that is binned post-hoc for inference.  To address (2), we have to use binned inference since we do not know the unbinned analog of Eq.~\ref{eq:chi2}.  Lastly, we will compare asymptotic and numerical uncertainties to answer (3).  Reference~\cite{Andreassen:2019cjw} found that binning before or after unfolding gave similar accuracy, but a study of precision for downstream parameter estimation has not yet been addressed.   
Since the detector distortions are treated unbinned, we hypothesized that unbinned unfolding could be more precise than traditional, binned unfolding.  Furthermore, we expected that using correlations in the test statistic would lead to more precise inference than ignoring correlations in the test statistic.  Lastly, we hypothesized that the asymptotic formulae would not give correct uncertainties when correlations are ignored (and correct ones when they are included when possible).  
For a series of Gaussian examples, we will confront each of these hypotheses with empirical evidence in order to address our main questions above.

The remainder of this paper is organized as follows. In Sec.~\ref{sec:methods}, we detail the methodology of unbinned unfolding and inference.  Section~\ref{sec:setup} outlines the experimental setup, including the construction of simulated datasets. Numerical results for one and many dimensions are presented in Sec.~\ref{sec:results}.  The paper ends with conclusions and outlook in Sec.~\ref{sec:conclusion}.


\section{Methodology}
\label{sec:methods}


\subsection{Unbinned Unfolding}

We studied unbinned unfolding using the OmniFold~\cite{Andreassen:2019cjw,Andreassen:2021zzk} method, as it is the one that has been used in most recent unbinned measurements.  The task is to find the true distribution given observed data from a detector that includes non-negligible resolution or smearing effects.  An essential requirement is to have an accurate detector simulation that maps the generated particle level observables (gen particle) to the output of the detector (sim particle).   An a-priori estimate of the true distribution is used to generate the gen-particle distribution in the Monte Carlo.  The a-priori estimate is necessary because the true distribution is what we would like to measure using the data and it is either not known in advance, or the goal is to improve upon prior measurements, which are the basis for the a-priori estimate, with the data that are the input to the unfolding. The OmniFold method finds the Monte Carlo event weights necessary to reweight the gen-particle distribution, from the a-priori estimate, to the true distribution using an iterative process.

Each OmniFold iteration consists of two steps.  The first step is to find the Monte Carlo event (``pull'') weights that will make the sim-particle distribution match the detector data.  The desired unfolding output is a function that gives an event weight from gen-particle inputs, not sim-particle inputs.  The second step of OmniFold provides this by finding Monte Carlo event (``push'') weights that will make the gen-particle distribution match the pull-weighted gen-particle distribution.  The push-weighted Monte Carlo sim-particle distribution can then be compared to the detector data distribution.  If they agree, the task is complete.  If they disagree, the next iteration repeats the procedure starting with the push-weighted Monte Carlo sample, where the product of the push weights of all previous iterations are retained as event weights.  The step weights in each iteration become closer to one as the process converges, usually after about five iterations.  The final output is the weighted generation, i.e. particle level Monte Carlo sample, where each event is weighted by the product of the push weights for all of the iterations.  


A detailed discussion of unbinned unfolding and the optimization of hyperparameters, such as the number of iterations, can be found in reference~\cite{Canelli:2025ybb}.


\subsection{Unbinned Inference}

Inference of model parameters from unbinned data is commonly done using a maximum likelihood fit.  Asymptotic uncertainties on the model parameters are determined from finding the parameter values that give a change in log likelihood of 0.5, corresponding to one standard deviation in the Gaussian approximation (e.g. the likelihood is locally quadratic near the maximum).  Methods exist~\cite{Langenbruch:2019nwe} for taking into account event weights in the evaluation of the asymptotic uncertainties if the weights are statistically independent.
Reference~\cite{Langenbruch:2019nwe} suggests a method for taking into account weight correlations if they can be parameterized.  For our case, there is no simple parameterization, but we leave further explorations of related strategies to future work.
In our study, we fit the output of the OmniFold unfolding, which is the weighted Monte Carlo gen-particle distribution.  The OmniFold event weights are not statistically independent.  We study how these weight correlations impact the evaluation of the asymptotic uncertainties from the maximum likelihood fit.

Statistically valid inference uncertainties, even when analyzing unfolding output events with weight correlations, can be obtained numerically, either by using the bootstrap method \cite{efron1979}, or from statistically independent pseudo datasets.  In the work presented here, we use the numerical evaluation of the inference uncertainties as a valid reference for our studies of the asymptotic inference uncertainties.  The numerical uncertainty is defined as the root-mean-square (RMS) of the distribution of inference results from either a collection of statistically independent pseudo datasets or a set of bootstrap samples derived from the dataset being unfolded and analyzed.  We use "RMS" as a useful shorthand for the numerical uncertainty in the text and figures below.


\subsection{Implementation}

We consider both a one dimensional and multi-dimensional inference task described in the next section.  For the one-dimensional study, we use kernel density estimation (KDE) with a uniform kernel over a fixed radius to estimate the probability densities that are used within OmniFold.  The estimated probability density function (PDF) $p$ at point $x$ in the feature space is proportional to the number of points within $\delta x$ of $x$.  The pull weight in step 1 for a Monte Carlo event at $x_s$ is $w_1 = p_d(x_s)/p_s(x_s)$, where $p_d$ and $p_s$ are the PDFs for the data and sim-particle Monte Carlo, respectively.  Likewise, the push weight for step 2 is $w_2 = p_g(x_g,w_1)/p_g(x_g)$, where $p_g(x_g,w_1)$ is the gen-particle PDF evaluated with the weights $w_1$ from step 1. 

In higher dimensions, we instead train neural network binary classifiers to find the weight functions used in OmniFold.  In step 1, a classifier is trained to distinguish data from sim-particle Monte Carlo and the pull weight is $w_1 = c_1(x_s)/(1-c_1(x_s))$, where $c_1(x_s)$ is the classifier output, which is the probability that an event at $x_s$ is from the data distribution.  In step 2, another classifier is trained to distinguish gen-particle Monte Carlo weighted by $w_1$ from gen-particle Monte Carlo and the push weight is $w_2 = c_2(x_g)/(1-c_2(x_g))$, where $c_2(x_g)$ is the classifier output.


\section{Setup}
\label{sec:setup}

To demonstrate parameter estimation in addition to unfolding, we need data generated from a model with a known parametric family.  In order to run enough (pseudo)experiments to test the coverage of the inference (in addition to research and development), we also need to be able to generate many examples from the model.  Therefore, we focus on Gaussian examples.  Many physics problems of interest share many features with this setup; at the same time, we expect the sign and magnitude of the features we observe to be problem-specific.


\subsection{Data}

In our one-dimensional studies, the true model is a Gaussian with parameters $\mu_{\rm true}=0.2$ and $\sigma^2_{\rm true}=0.81$.  The parameter values used in the multidimensional studies are given in Appendix~\ref{app:nd-model-par-vals}.  10,000 events are generated for the ``truth'' dataset.  A larger Monte Carlo (MC) sample of size 100,000 events is also generated from a Gaussian with mean 0.0 and variance 1.0.  In traditional unfolding, this MC dataset would be the one to derive the response matrix.  It also serves as the initialization for the unfolded result.  Both the true and `particle-level' MC sample are passed through the same detector distortions: a Gaussian smearing with mean zero and various resolutions (i.e. detector distortions) ranging from 0 to 0.75. For each parameter set, datasets are generated and binned into a fixed number of bins (15 bins spanning the range $[-3,4]$).  The true mean is slightly positive, so the range included in the binning is larger on the positive side.  In the unbinned cases, the result is unbinned and presented as a histogram for illustration and comparison purposes.

\begin{figure}[ht!]
   \begin{center}
\includegraphics[width=0.95\linewidth]{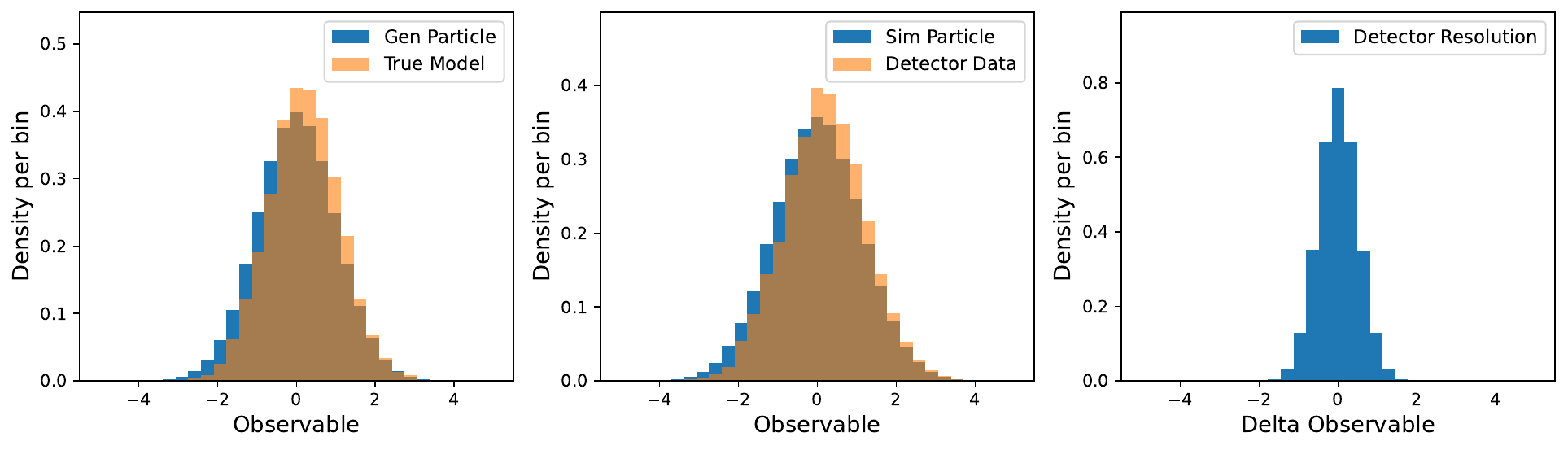}
   \caption{
      Histograms showing the datasets from the one-dimensional study.  The distributions on the left show the true distribution of the observable compared with the gen-particle distribution of the Monte Carlo.  The center histograms show a sample detector-level distribution of the observable compared with the sim-particle distribution of the Monte Carlo.  The right histogram shows the resolution function.  For these histograms, we used 31 bins in the range [-5, 5].
   }
   \label{fig:distributions-1d}
   \end{center}
\end{figure}

For each parameter set that we study, we generate 500 statistically independent pseudo datasets where the number of events in each dataset is chosen from a Poisson distribution with a mean of 10,000.  Figure~\ref{fig:distributions-1d} shows example distributions.  In a real application, where there is only one dataset being analyzed, the bootstrapping method~\cite{efron1979} could be used to determine the statistical uncertainties numerically.  We have verified that bootstrapping and independent pseudo datasets give the same results for the statistical uncertainties, as evaluated by the RMS of the inference results of 500 datasets.

\subsection{Neural Networks}

We use the Tensorflow 2.17.0~\cite{tensorflow2015-whitepaper} and Keras 3.4.1 frameworks to implement and train our Neural Networks (NNs).  The NN consists of three fully connected layers with 50 nodes per layer.  In our nominal results, we use a modified version of the {\tt gelu}~\cite{hendrycks2023gaussianerrorlinearunits} activation function, though we have also explored using {\tt relu} activation.  The {\tt gelu} function is modified as
\[
{\rm GELU}(x) \approx 0.5 \ x \left( 1 + \tanh\left[ \beta \left(x+0.044715 \, x^3\right)\right] \right) \ ,
\]
where $\beta$ is a parameter that tunes the distance scale of the transition from flat to a slope of 1.  With $\beta=1$, the function is the original {\tt gelu}, which gives a high degree of regularization and a larger distance scale for the output event weight correlations.  With $\beta=4$, the regularization is more modest and the correlations are closer to the output with {\tt relu}, which is the asymptotic case for large $\beta$.  More details on our choice of activation function are given in Appendix~\ref{app:activation}. 
The NNs are trained using the Adam optimizer~\cite{kingma2017adammethodstochasticoptimization} with a learning rate of $5 \times 10^{-4}$.  The loss function is binary cross entropy.  Data are equally divided into train and test samples for monitoring.  The NNs are trained for 50 epochs with a batch size of $10^4$.
We remove the output dependence on the random initial NN model weights, which are set with the {\tt GolrotUniform} method~\cite{pmlr-v9-glorot10a}, by averaging the final event weights of 10 runs of the full OmniFold procedure, each with different random initial model weights.  The effects of this ensembling~\cite{Acosta:2025lsu} are described in more detail in Appendix~\ref{app:ensembling}. 

These hyperparameter values were chosen to give stable results with reasonable convergence.  When exploring the hyperparameter space, we monitored several aspects of the unfolding, including:
\begin{itemize}
    \item The mean and RMS of the step 1 and step 2 weight distributions for each OmniFold iteration.  Satisfactory convergence is indicated by a mean weight of 1 and a narrow weight distribution.
    \item The residuals of the unfolded distribution, compared to the known true distribution after each iteration for both the gen-particle and sim-particle distributions in all feature dimensions.  
    \item The mean and width of the distribution of the difference between the OmniFold event weight and the known true event weight.  Since the multivariate normal parameters are known in our study for both the Monte Carlo and true distribution, the true event weight can be calculated for each event.  For the 1D study, this can be visualized as a simple graph. Example graphs of the event weight function can be found in Appendix~\ref{app:activation}.
\end{itemize}
Examples of the monitoring output can be found in the code repository for this paper. 

As an alternative to NNs in the one-dimensional studies, we also use KDEs with a fixed kernel of fixed radius 0.1.


\section{Numerical Results}

We start with the one-dimensional example and first establish a binned baseline.  Then, we consider the unbinned case (also in one dimension), by first characterizing the size of the correlations.  Lastly, we explore the higher-dimensional inference task (which has no binned baseline).


\label{sec:results}
\subsection{Binned Baseline}
\subsubsection{Binned Unfolding with IBU and Inference}

For comparison with unbinned approaches, we establish a baseline in which the data are binned from the start.
We perform Iterative Bayesian Unfolding (IBU)~\cite{DAgostini:1994fjx} (also known as Lucy-Richardson deconvolution~\cite{Richardson:72,1974AJ.....79..745L}) on the binned data, which is similar to OmniFold in the binned limit.
Both IBU and OmniFold are iterative processes that converge to the maximum likelihood estimate of the particle-level distribution given the detector-level data (see the Appendix of~\cite{Andreassen:2019cjw}).
The unfolded binned distribution is fit to a binned Gaussian using a $\chi^2$ statistic.
In our inference, two variants are considered: one that employs the full covariance matrix computed from 500 bootstrap replicas of the unfolded data, and another that uses only the diagonal elements of the covariance matrix.
The covariance is estimated numerically from the ensemble of bootstrapped histograms.
Parameter estimation is performed by minimizing the $\chi^2$, and the uncertainties are determined either by the asymptotic change in $\chi^2$ (with $\Delta\chi^2=1$) or numerically from the spread (standard deviation or RMS) of the inferred parameters over the bootstrap ensemble.

This binned baseline procedure provides a benchmark against which the performance of the unbinned unfolding (followed by either post-hoc binning for inference, or unbinned inference) can be compared.
For each parameter, two sets of asymptotic analyses are performed:
one obtained from the fit using the full covariance matrix (``cov'') and one using only the diagonal approximation (``cov\_diag'').
We then calculate the mean asymptotic error and its standard error (SEM) over the 500 replicas.
In parallel, the numerical uncertainty is obtained from the standard deviation of the best-fit parameter values across these replicas.

\begin{figure}
    \centering
    
    \subfloat[]{\includegraphics[width=0.43\linewidth]{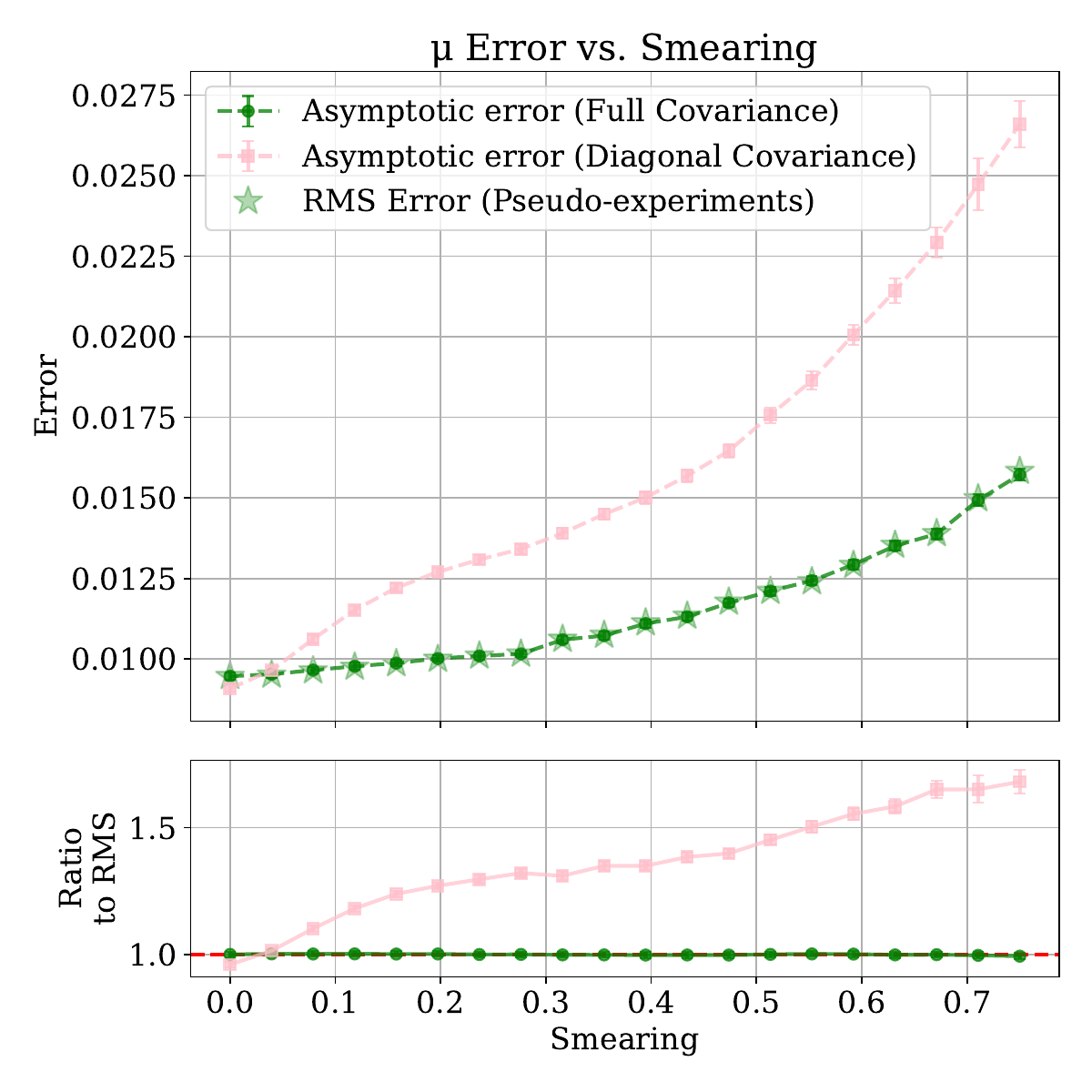}\label{fig:mu_error_plot_with_errorbars}}\quad\subfloat[]{\includegraphics[width=0.45\linewidth]{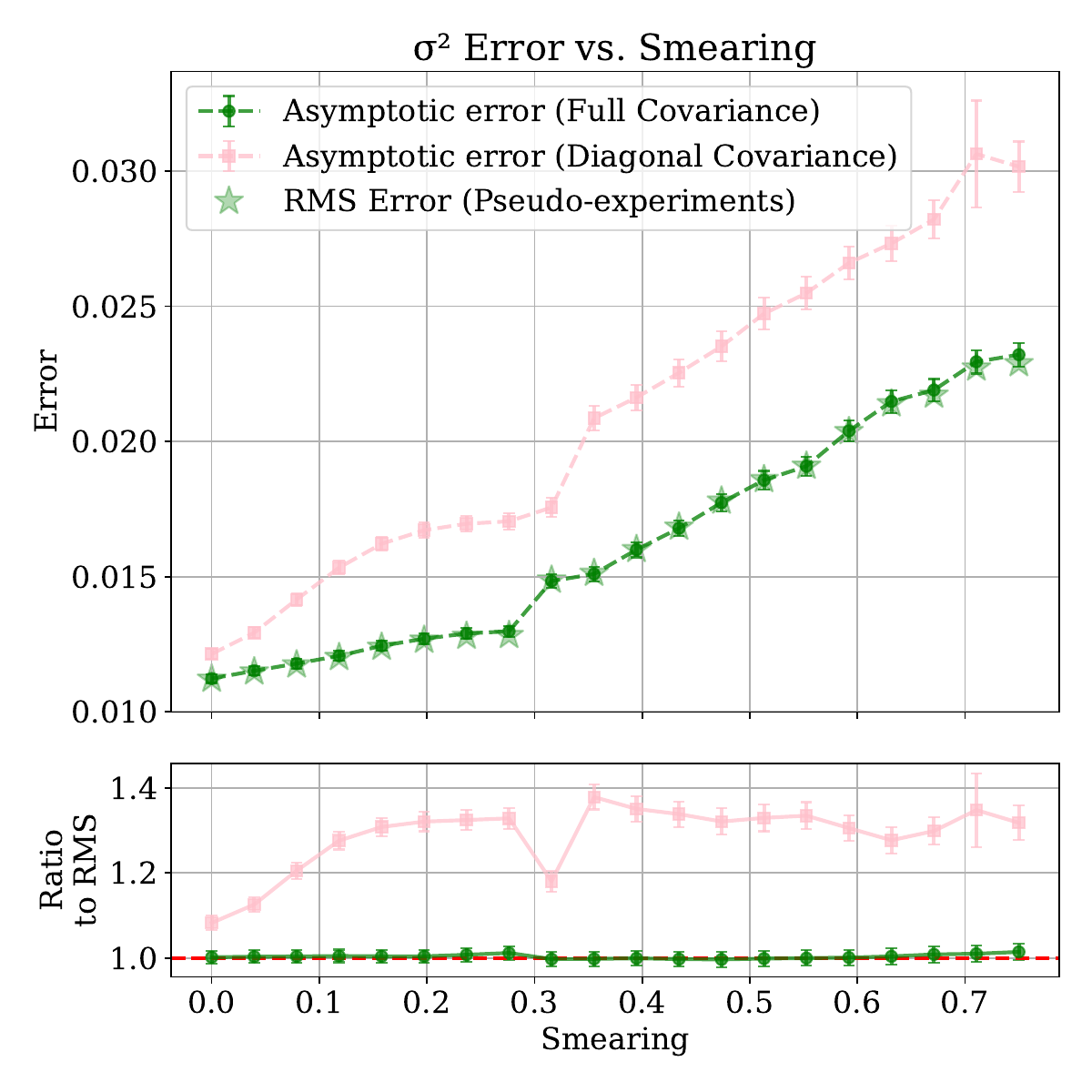}\label{fig:var_error_plot_with_errorbars}}
    \caption{ Mean asymptotic error versus detector smearing for the (a) Gaussian mean, $\mu$, and (b) the variance $\sigma^2$ obtained from the full covariance analysis (green circles) and the diagonal covariance approximation (pink squares).
    The green stars represent the standard deviation computed from the spread of the best-fit values over 500 bootstrap replicas.
    Note that the numerical uncertainty agrees well with the asymptotic error from the full covariance fit while the diagonal approximation consistently overestimates the uncertainty at larger smearing values. 
    }
    
    \label{fig:uncertsfullybinned}
\end{figure}

First, we study the uncertainties in Figures~\ref{fig:mu_error_plot_with_errorbars} and~\ref{fig:var_error_plot_with_errorbars} for the fitted mean ($\mu$) and variance ($\sigma^2$), respectively.
As the detector resolution deteriorates, the overall uncertainty increases, as expected.
Notably, the numerical uncertainty matches the asymptotic error using the full covariance matrix but is consistently smaller than the asymptotic error computed using the diagonal covariance matrix, particularly for larger smearing values.
This indicates that the standard asymptotic error estimation using only the diagonal values of the covariance matrix overestimates the true confidence intervals when the unfolding-induced correlations become significant.
The cause of the jump in the graphs of~\ref{fig:var_error_plot_with_errorbars} is not known, but this does not affect our conclusions.
These findings underscore the importance of incorporating the full covariance structure in the inference process when computing asymptotic uncertainties.

\begin{figure}
    \centering
    
    \subfloat[]{\includegraphics[width=0.43\linewidth]{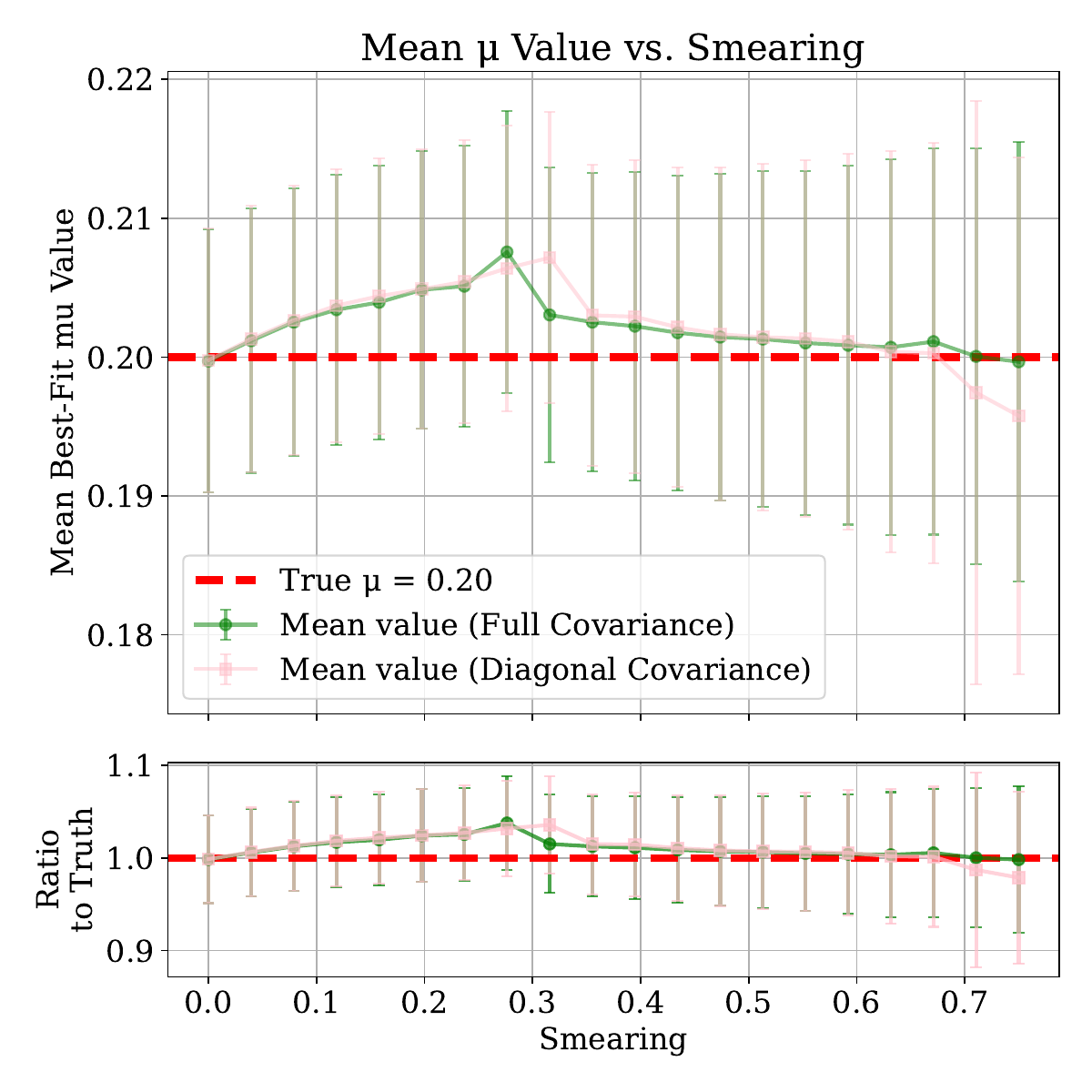}\label{fig:mu_mean_values_with_errorbars}}\quad\subfloat[]{\includegraphics[width=0.43\linewidth]{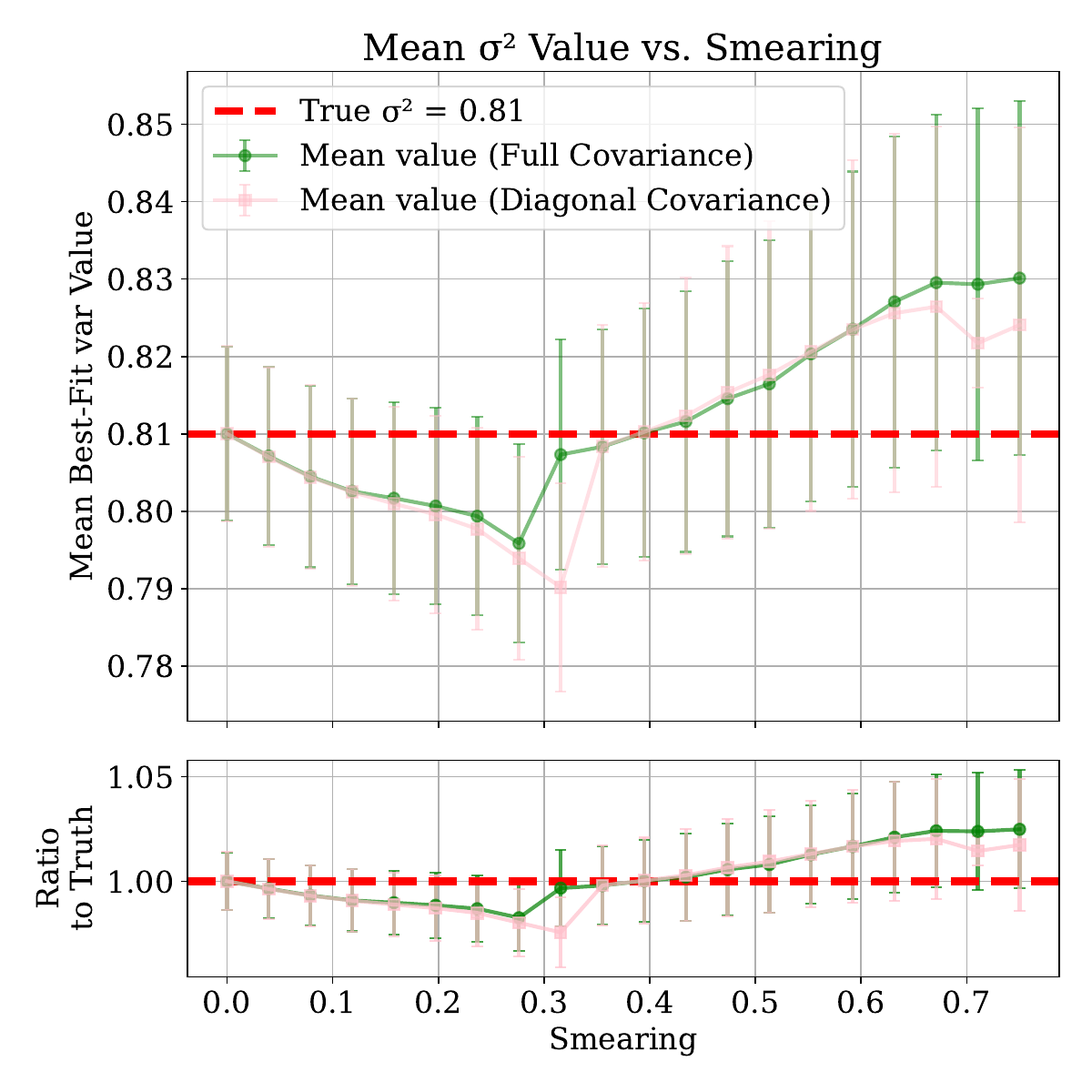}\label{fig:var_mean_values_with_errorbars}}
    \caption{The mean best-fit value of (a) $\mu$ and (b) $\sigma^2$ is shown as a function of the detector smearing, with error bars indicating the standard error of the mean (SEM) over 500 bootstrap test datasets.
    Horizontal dashed red lines mark the true values of $\mu=0.2$ and $\sigma^2=0.81$.
    Both fit methods (full covariance and diagonal approximation) yield similar central values. 
    }
\end{figure}

Next, we quantify the bias in Figures~\ref{fig:mu_mean_values_with_errorbars} and~\ref{fig:var_mean_values_with_errorbars} for the evolution of the mean best-fit values of $\mu$ and $\sigma^2$, respectively. In these plots, the data points represent the average best-fit value at a given smearing, with vertical error bars corresponding to the SEM computed from the 500 replicas. In addition, a horizontal dashed red line marks the true parameter value.
For both $\mu$ (Figure~\ref{fig:mu_mean_values_with_errorbars}), and $\sigma^2$ (Figure~\ref{fig:var_mean_values_with_errorbars}), the estimates computed both with and without using off--diagonal elements of the covariance matrix are similar.  In particular, all of the methods show nearly the same bias as a function of smearing, which is covered by the uncertainty.  The truth spectra are the same for each smearing and thus the uncertainties are very correlated between points.

These results provide an important baseline: they demonstrate that with IBU and conventional binned inference, the choice of covariance treatment can affect the inferred uncertainties.
In the subsequent sections, we compare these findings with those obtained from the unbinned unfolding approach.

\subsection{Event-by-event Correlations}

For each parameter set, 500 statistically independent test datasets are unfolded without binning using a fixed Monte Carlo sample.  Each Monte Carlo event $i$ in the output of unfolding dataset $j$ is assigned a weight $w_{ij}$ by the unfolding algorithm.  This allows us to quantify the degree to which the event weights for a pair of Monte Carlo events are correlated in the unfolding.  

\begin{figure}[ht!]
   \begin{center}
\includegraphics[width=0.95\linewidth]{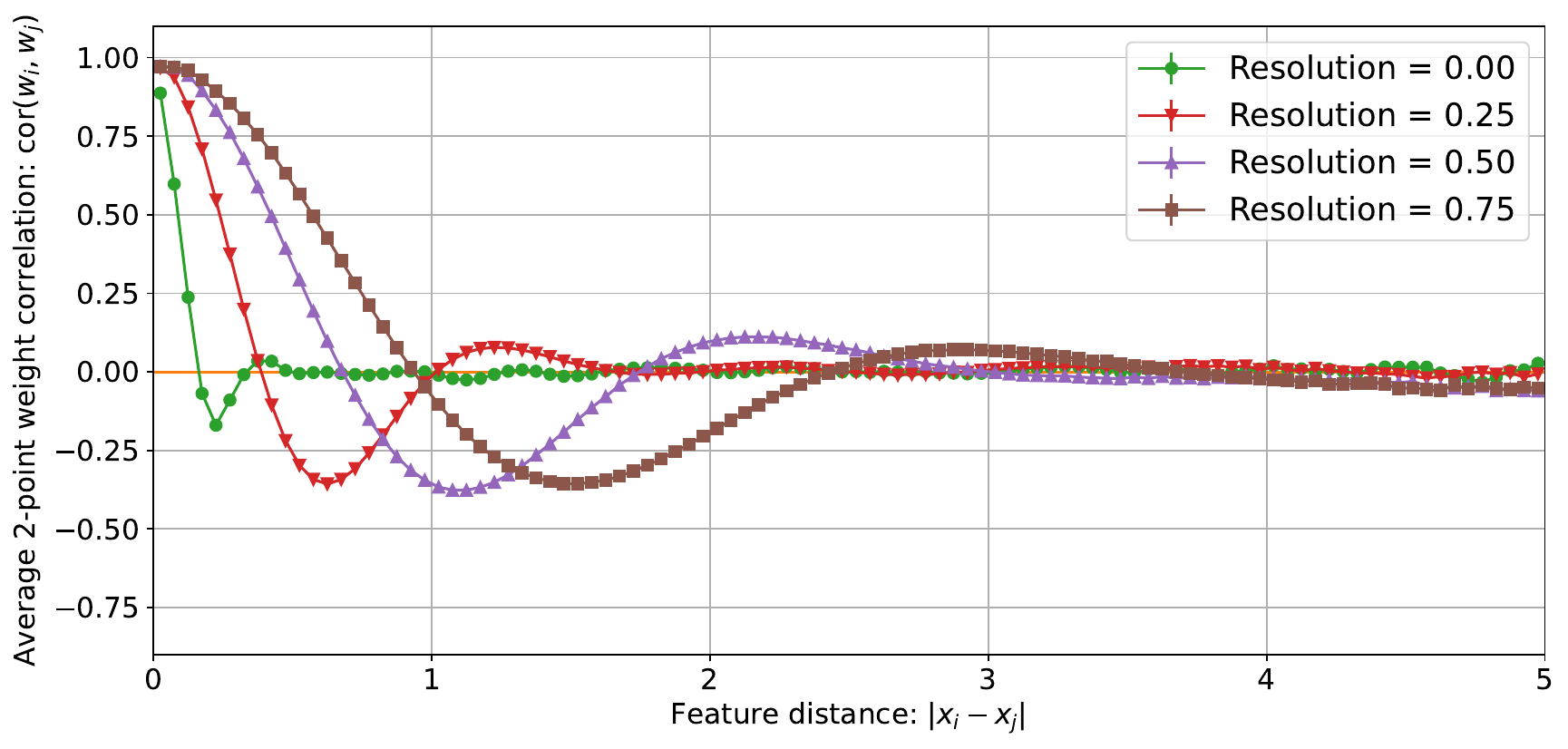}
\includegraphics[width=0.95\linewidth]{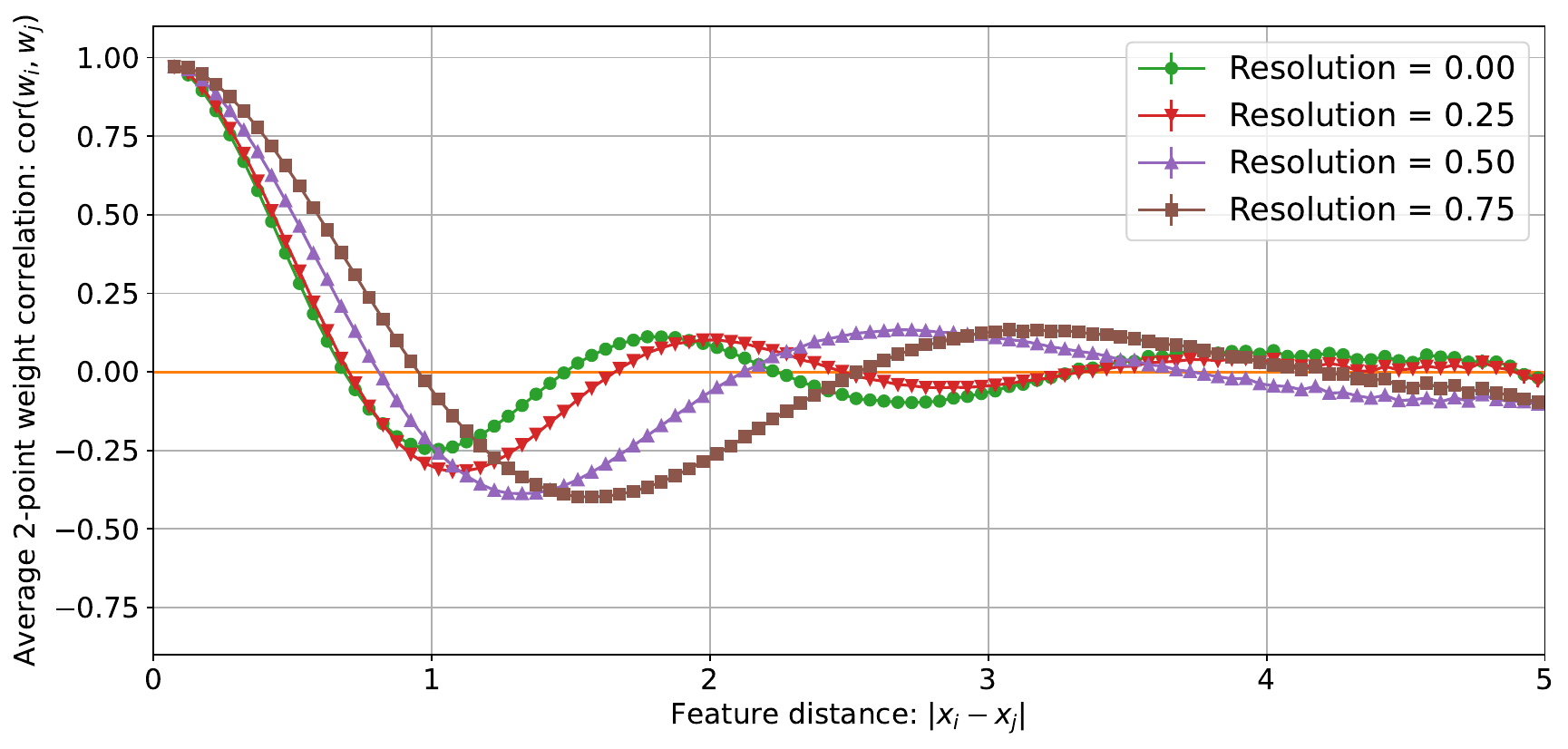}
   \caption{
      Average weight correlation between two events as a function of the absolute distance between the events in the observable.  The four curves show different values for the detector resolution.  The top plots show unfolding with the KDE approach within OmniFold, while the bottom plots show results from using NNs within OmniFold.  The error bars show the RMS of the correlation values.
   }
   \label{fig:weight-correlation-vs-distance-1d}
   \end{center}
\end{figure}

Figure~\ref{fig:weight-correlation-vs-distance-1d} shows the average weight correlation between pairs of events as a function of the distance between the pair of events in the observable for both the KDE and NN approaches.  The closer the events are to each other, the stronger the weights are correlated, with a correlation of 1 at a distance of zero by construction.  For perfect resolution, the correlation quickly dies off as the distance between the events grows.  In principle, in this case, there should be no correlations between events and so any residual correlation is due to smoothing in the KDE or NN.  
The implicit regularization within the KDE and NN is different.  For example, the KDE range parameter ($\delta x$) limits the length scale of correlations, in the absence of detector smearing, while the NN has no explicit length scale built into it.
As the detector resolution gets worse, the range of the correlation increases.  The correlations themselves show a wave-like pattern as nearby events are correlated and then anti-correlated beyond some point in order to preserve the total number of events.  

\begin{figure}[ht!]
   \begin{center}
\includegraphics[width=0.95\linewidth]{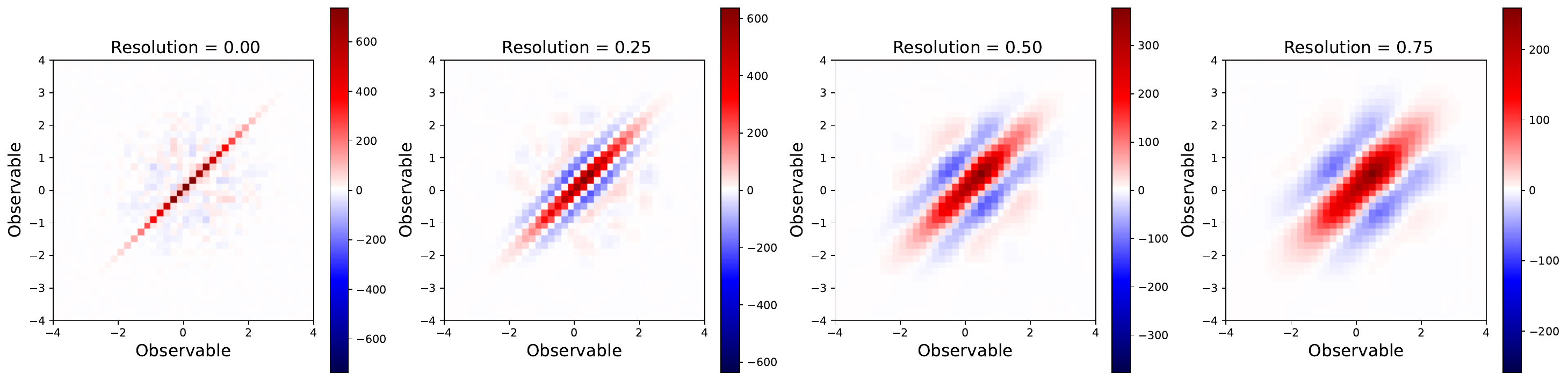}
\includegraphics[width=0.95\linewidth]{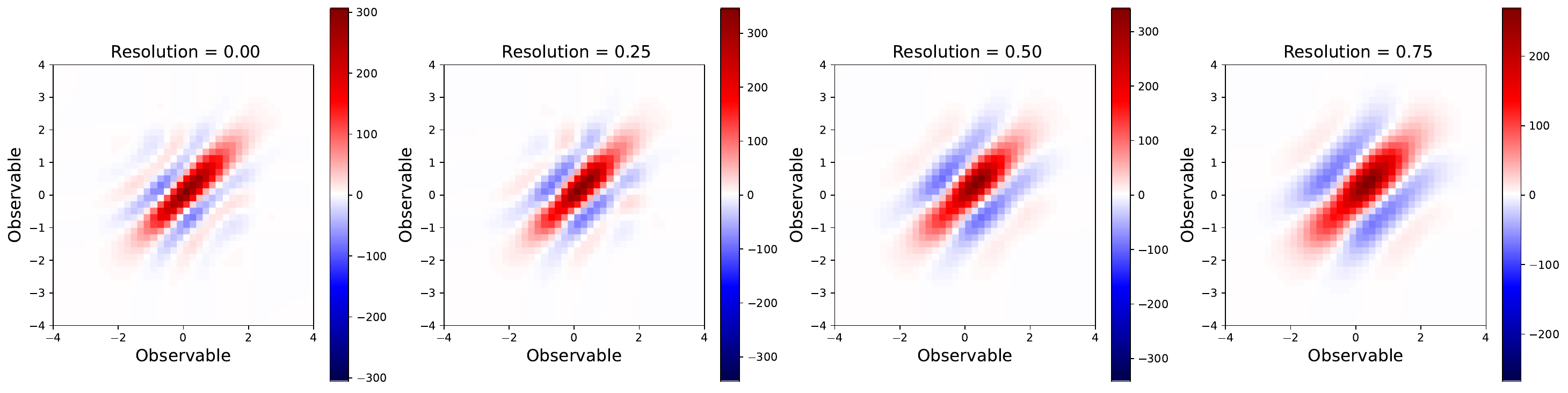}
   \caption{
      Covariance matrices of a histogram of the unfolding output for four detector resolution values (0, 0.25, 0.50, 0.75).
      The top plots show unfolding with the KDE approach within OmniFold, while the bottom plots show results from using NNs within OmniFold.  The histogram has 40 bins in the range [-4, 4].
   }
   \label{fig:weight-histogram-covariance-1d}
   \end{center}
\end{figure}

Another way to visualize the correlations is to bin the data and construct covariance matrices.  Figure~\ref{fig:weight-histogram-covariance-1d} shows these matrices using histograms (40 bins from -4 to 4 in the observable) of the unfolding output, where the covariance is calculated numerically from the 500 test datasets for each resolution value.  The results are consistent with the unbinned evaluation of the weight correlations, shown in Figure~\ref{fig:weight-correlation-vs-distance-1d}, and additionally show that the distance scale of the correlations does not depend strongly on the observable itself.

\subsection{Parameter Estimation in 1D with a KDE}

To investigate the impact of the event weight correlations in the unbinned unfolding output, we compare three ways of performing the model parameter inference:
\begin{itemize}
    \item An unbinned ML fit of the unbinned unfolding output, which is the weighted Monte Carlo sample.
    \item A binned $\chi^2$ fit of the unfolding output that uses the full covariance matrix for the histogram bins.
    \item A binned $\chi^2$ fit of the unfolding output that uses only the diagonal elements of the covariance matrix for the histogram bins.
\end{itemize}
For the binned $\chi^2$ fits, the covariance matrix of the histogram bins is evaluated numerically from a set of histograms made from the 500 test datasets.  
The inference uncertainty is determined for each model parameter from the RMS of the inference results for the 500 test datasets.  We consider this to be the most correct estimate of the inference uncertainty.  We compare this with the asymptotic uncertainties from the fits, determined from the change in the model parameter that gives a change of 0.5 in the log likelihood for the ML fit or a change of 1 in the $\chi^2$.  We also compare the unbinned unfolding with the results of performing a binned unfolding using the Iterative Bayesian Unfolding technique.  For all binned results, the Gaussian model predictions are integrated over the bins in order to have histogram templates that are properly parameterized.

\begin{figure}[ht!]
   \begin{center}
\includegraphics[width=0.95\linewidth]{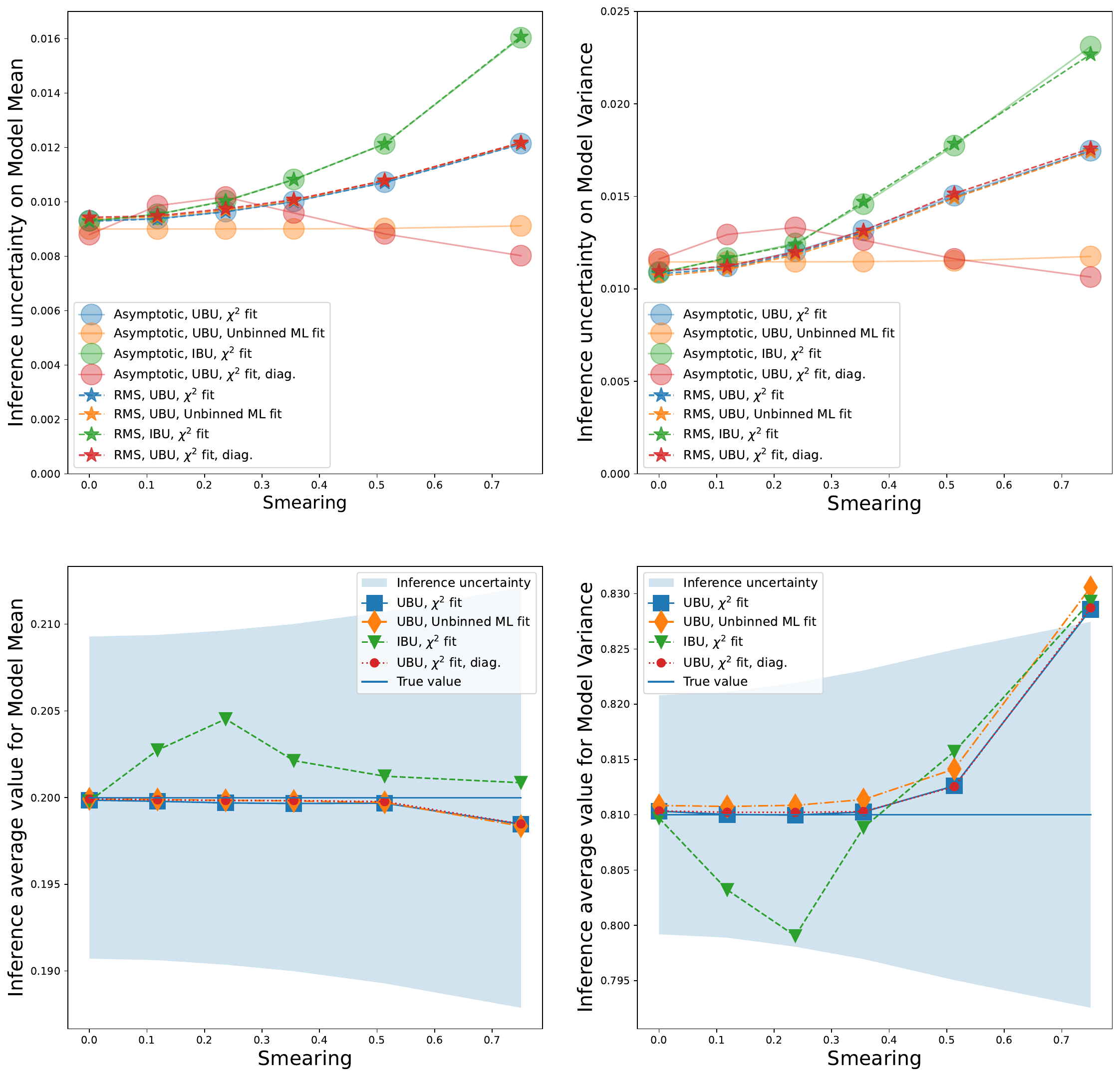}
   \caption{
      Inference uncertainty (top) and bias (bottom) on the model mean (left) and variance (right) as a function of the detector resolution.  Unbinned unfolding (UBU) is compared with the results of binned Iterative Bayesian Unfolding (IBU).  The inference is done three ways: an unbinned ML fit, a binned $\chi^2$ fit using the full covariance matrix for the fit histogram, and a binned $\chi^2$ fit using only the diagonal elements of the covariance matrix for the histogram. 
      The blue band centered on the true value shows $\pm$ the RMS inference uncertainty from the unbinned unfolding unbinned ML fit to provide a scale reference.
   }
   \label{fig:uncertainty-and-bias-vs-resolution}
   \end{center}
\end{figure}

Figure~\ref{fig:uncertainty-and-bias-vs-resolution} shows the inference uncertainty and bias as a function of the detector resolution.  We draw the following conclusions from the results shown:
\begin{itemize}
    \item The asymptotic uncertainties in the unbinned ML fit of the unbinned unfolding are too low, do not depend on the detector resolution, and agree with the correct uncertainty only for the perfect detector resolution case. This can be seen in the top plots, noting that the orange circles do not depend on the smearing and they disagree with the orange stars (hidden under the red stars) for all levels of smearing except no smearing.  
    \item The asymptotic uncertainties in the binned $\chi^2$ fits agree with the correct uncertainty, as evaluated numerically from the RMS of the inference distribution, when the full covariance matrix is used.  Pictorially, blue circles agree with blue stars (hidden under the red stars) and the green circles agree with the green stars. 
    \item When the off-diagonal elements of the covariance matrix are set to zero in the $\chi^2$ fit, the asymptotic uncertainties are incorrect. This corresponds to the difference between the red circles and red stars, is the binned analog of the first point, and was observed in the fully binned case in Fig.~\ref{fig:uncertsfullybinned}).
    \item From the lower plots, we conclude that the bias in the unbinned ML fit is the same as the bias in the $\chi^2$ fit.
    \item The inference uncertainty and bias for the unbinned unfolding is better than the for the binned unfolding with IBU
    for the binning and method hyperparameter choices that produced the results shown in Figure 6.
    This suggests that some useful information may be lost when the data are binned before unfolding. 
\end{itemize}

The results from the $\chi^2$ fits are as expected.  When correlations from the unfolding are properly included by using the full covariance matrix for the fit histogram, the asymptotic uncertainties are correct.  When these correlations are ignored, the asymptotic uncertainties are incorrect.  Correlations in the unbinned unfolding are negligible when the detector resolution is perfect and, in this case, the asymptotic uncertainties in the unbinned ML fit are correct because the event weights from the unfolding are uncorrelated.

\subsection{Parameter Estimation in 1D with a NN}

\begin{figure}[ht!]
   \begin{center}
\includegraphics[width=0.95\linewidth]{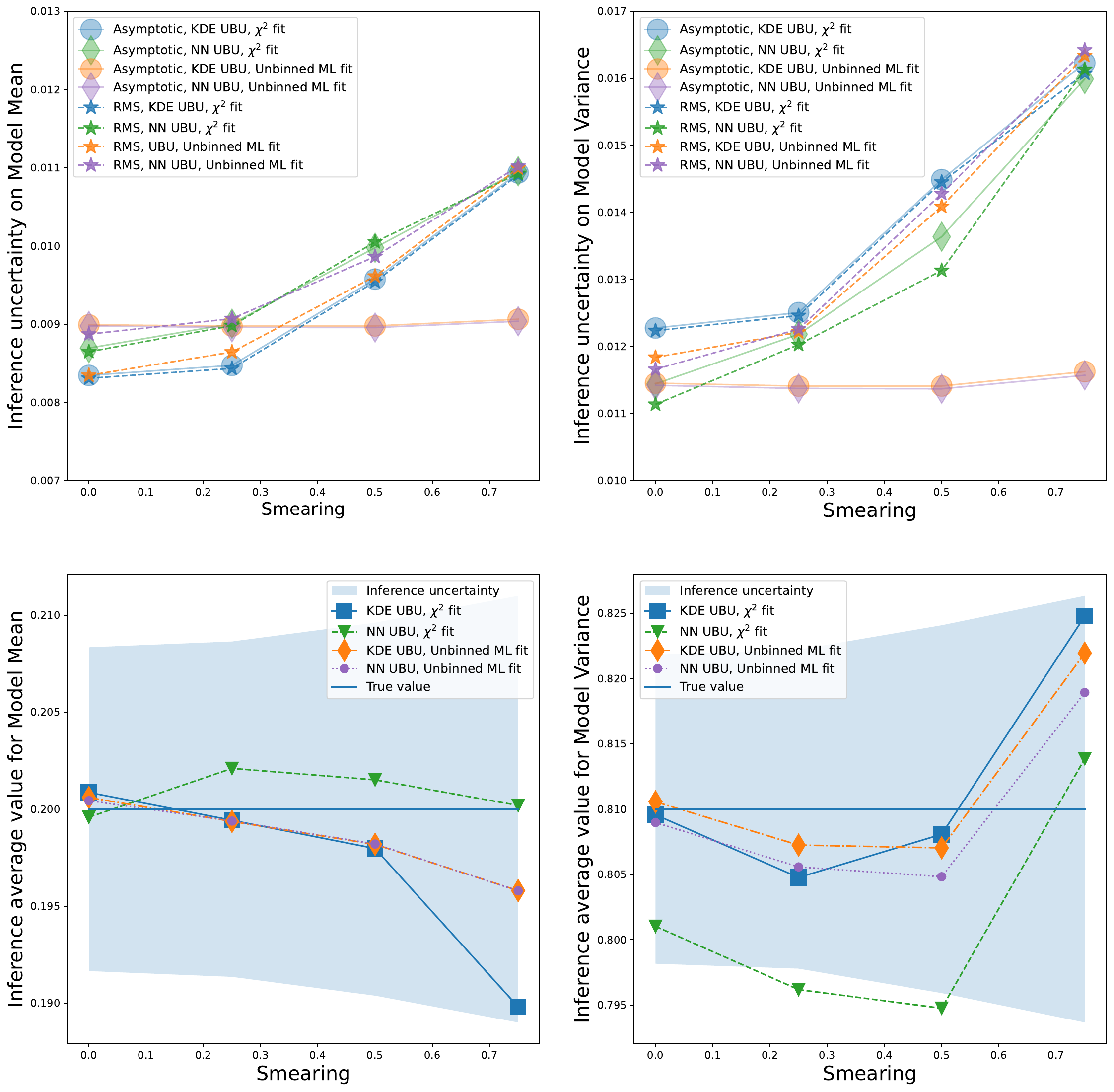}
   \caption{
      Inference uncertainty (top) and bias (bottom) on the model mean (left) and variance (right) as a function of the detector resolution.  Unbinned unfolding (UBU) is done two ways: with the KDE and with Neural Networks (NN) in OmniFold.  The inference is done two ways: an unbinned ML fit and a binned $\chi^2$ fit using the full covariance matrix for the fit histogram. 
      The blue band centered on the true value shows $\pm$ the RMS inference uncertainty from the unbinned unfolding unbinned ML fit to provide a scale reference.
   }
   \label{fig:uncertainty-and-bias-vs-resolution-nn}
   \end{center}
\end{figure}

Figure~\ref{fig:uncertainty-and-bias-vs-resolution-nn} has a similar setup as Fig.~\ref{fig:uncertainty-and-bias-vs-resolution}, but compares the KDE and NN approaches within OmniFold.  The trends with the NN are similar.

We found that the activation function can affect the degree of regularization in the NN outptut, which also affects the strength and distance scale of the weight correlations in the OmniFold output.  Interestingly, we found that the inference uncertainty is insensitive to these correlations.  This is discussed in more detail in Appendix~\ref{app:activation}.

\subsection{Parameter Estimation beyond 1D}

In dimensions higher than 2 or 3, unbinned unfolding becomes the only practical method.  We investigate higher-dimensional feature spaces using a multivariate normal distribution as our `physics' model.  The covariance parameters of the model are chosen such that the densities do not factorize in the variables spanning the feature space.  
The mean and sigma values were chosen to include a range of mean values within [-1,1], sigma values within [0.6, 1.5], and non-trivial off-diagonal correlation coefficients within [-0.7,0.7].  The Monte Carlo gen particle and pseudo data true values differ by small amounts, so that the unfolding task is non trivial.  The chosen values give an invertible model multivariate Gaussian covariance matrix.
We model the detector resolution with a Gaussian smearing that is separate and uncorrelated for each feature dimension.  The model and resolution parameters used are given in Appendix~\ref{app:nd-model-par-vals}.

Figure~\ref{fig:nn-uncertainties-vs-nd} compares the inference precision as estimated with the RMS of the 500 pseudo datasets with the asymptotic uncertainty from an unbinned ML fit, using Eqn.~\ref{eq:unbinned} for each model parameter for studies in 1D, 2D, 4D, and 6D feature space.  The detector resolution is set to nominal values that are kept the same when additional feature dimensions are added.  In each study, the RMS uncertainty is higher than the asymptotic uncertainty.  The RMS/asymptotic uncertainty ratio appears to be the roughly the same for all of the model parameters within each plot with the ratio ranging from 1.18 to 1.28.

\begin{figure}[ht!]
   \begin{center}
\includegraphics[width=0.95\linewidth]{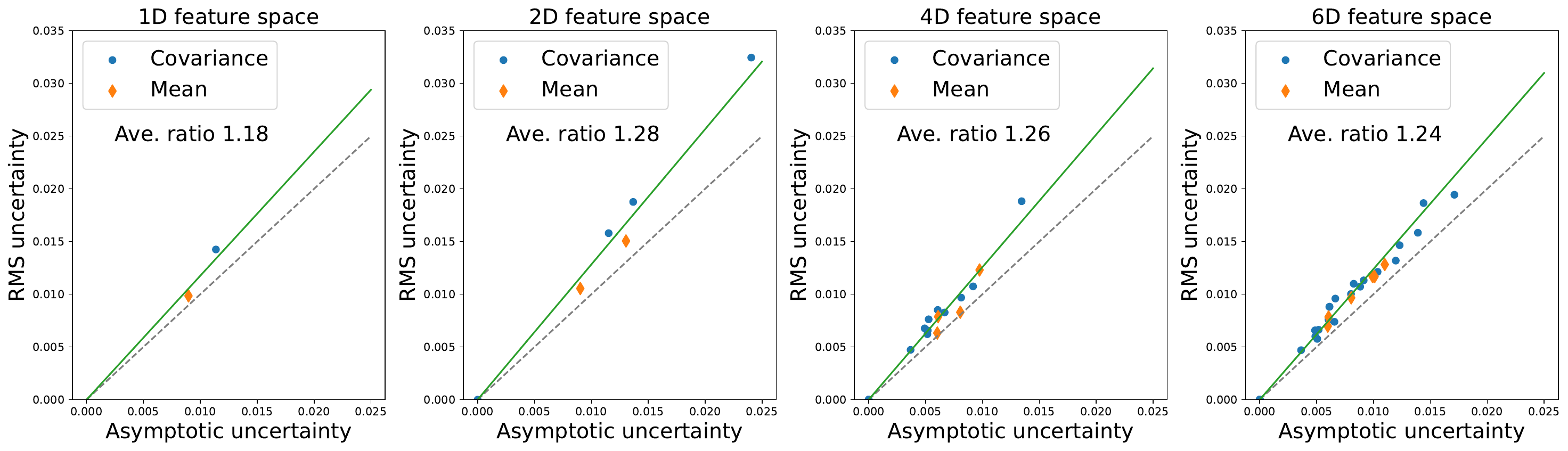}
   \caption{
      Scatter plots of the RMS vs average asymptotic inference uncertainty for unbinned unfolding using NNs in OmniFold in 1D, 2D, 4D, and 6D.  Each model parameter is represented by a point ($N$ means in $N$ dimensions and $N(N+1)/2$ covariance elements).  The vertical coordinate is the RMS of the fitted model parameter from 500 experiments, while the horizontal coordinate is the average asymptotic uncertainty from an unbinned ML fit.  The dashed gray line has a slope of 1.  The slope of the solid green line is set to the average of the RMS / asymptotic uncertainty ratio. 
   }
   \label{fig:nn-uncertainties-vs-nd}
   \end{center}
\end{figure}

\begin{figure}[ht!]
   \begin{center}
\includegraphics[width=0.95\linewidth]{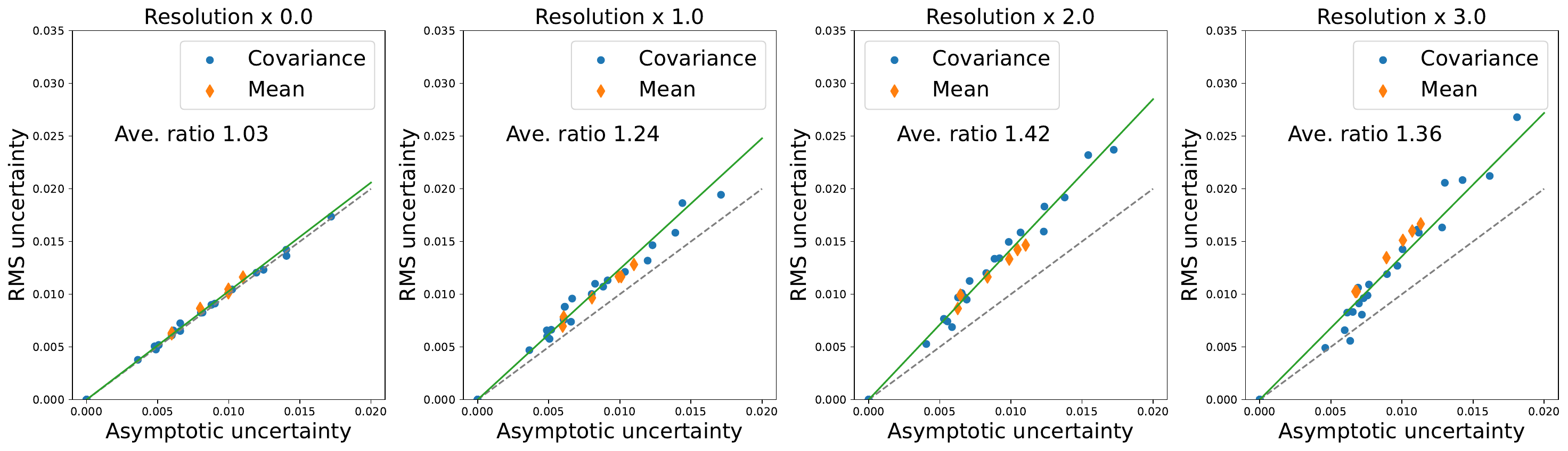}
   \caption{
      Scatter plots of the RMS vs average asymptotic inference uncertainty for unbinned unfolding using NNs in OmniFold in 6D with varying amounts of detector resolution smearing.  The detector resolution from the 6D study shown in Figure~\ref{fig:nn-uncertainties-vs-nd} is scaled by the following factors: 0, 1.0, 2.0, and 3.0.
      Each model parameter is represented by a point.  The vertical coordinate is the RMS of the fitted model parameter from 500 experiments, while the horizontal coordinate is the average asymptotic uncertainty from an unbinned ML fit.  The dashed gray line has a slope of 1.  The slope of the solid green line is set to the average of the RMS / asymptotic uncertainty ratio. 
   }
   \label{fig:nn-uncertainties-6d-res-var}
   \end{center}
\end{figure}

Figure~\ref{fig:nn-uncertainties-6d-res-var} compares four studies in the 6D feature space with varying detector resolution.  The results are qualitatively similar to the 1D case.  The RMS/asymptotic uncertainty ratio is consistent with 1 for perfect detector resolution and increases gradually as the detector smearing increases.

\clearpage

\section{Conclusions and Outlook}
\label{sec:conclusion}

Simulation-based inference is enabling a new class of analyses that make use of all the available information without first (lossily) compressing the data into a small number of summary statistics that are discretized via histograms.  New tools have enabled unbinned and high-dimensional inference of parameters and unbinned and high-dimensional inference of differential cross sections.  In this paper, we explore the fusion of these two approaches: unbinned parameter estimation from unbinned unfolded results.

We used a Gaussian setup, where the simulation/forward model of the data is known analytically and pseudodata were unfolded using the OmniFold unbinned unfolding method.  We identified a number of critical observations: (1) there can be an advantage of unbinned unfolding for parameter estimation even if the inference is performed using histograms.  This matches with our intuition that it is best to reduce the information as late as possible in the statistical analysis.  (2) Ignoring correlations -- as is currently required in the unbinned inference case -- seems to have little or no effect on the accuracy and precision of the fit as long as numerical methods are used for the statistical analysis.  This does not seem to be a formal consequence of the setup and may be a feature of the Gaussian example.  The motivating question of the entire study led us to (3).  Asymptotic formulae are not valid when correlations are ignored and the reported uncertainties can be significantly underestimated without computing them numerically.  

With a growing number of unbinned cross section measurements, it is critical to build a robust statistical and software suite for analyzing these data in the context of physics models.  Our main recommendation is to avoid asymptotic formulae for unbinned inference of unbinned measurements until the statistical formalism can be extended to include event-to-event correlations.  
Valid inference uncertainties can still be obtained, when analyzing the output of unbinned unfolding with event weight correlations, if they are evaluated numerically using the bootstrap method.
An important area of future research will be to directly study the interplay between machine learning regularization and the impact on downstream inference.  In this work, we have focused on the case where the `physics' model of the data is known; in the most general case, this model is not known explicitly and one must also use machine learning to approximate it from simulations.  Studying the end-to-end approach may reveal connections between the (implicit) regularization used by the machine learning models at both steps.  While there is still significant research required to optimally employ these methods, current results can already be used with care.  The scientific gains from these approaches hold immense potential for particle physics and beyond.

\section*{Code availability}
The code in this work can be found in: \url{https://github.com/owen234/unbinned-inference-paper}.

\section*{Acknowledgments}


We thank Jesse Thaler for useful discussions and feedback on the manuscript.
This material is based upon work supported by the National Science Foundation under Grant No. 2311666.



\appendix
\section{ Parameter values for the 2D, 4D, and 6D studies}
\label{app:nd-model-par-vals}


This appendix gives the parameter values used in the multidimensional studies.  The model parameters are the mean $\mu$, width $\sigma$, and correlation coefficients $\rho$ for the multivariate normal distribution.  Values are given for the Monte Carlo (MC) and the true values used to generate the pseudo data.
The detector resolution is $\sigma_{\rm det}$. 
\\

\noindent
Parameters for the 2D study. 
\[
\mu_{\rm MC} = 
\left(
\begin{array}{r}
0.0 \\ 1.0 \\  
\end{array}
\right) \ ,
\ \ \
\mu_{\rm true} = 
\left(
\begin{array}{r}
0.2 \\ 0.8 \\ 
\end{array}
\right) \ ,
\ \ \
\sigma_{\rm MC} =
\left(
\begin{array}{r}
1.0 \\ 1.5 \\ 
\end{array}
\right) \ ,
\ \ \
\sigma_{\rm true} =
\left(
\begin{array}{r}
0.9 \\ 1.3 \\ 
\end{array}
\right) \ ,
\ \ \ 
\sigma_{\rm det} = 
\left(
\begin{array}{r}
0.5 \\ 0.8 \\ 
\end{array}
\right)
\]
\[
  \rho_{\rm MC} =
  \left(
  \begin{array}{rr}
    1.0 & -0.6   \\
    -0.6 & 1.0   \\
  \end{array}
  \right) \ ,
\ \ \ \
  \rho_{\rm true} =
  \left(
  \begin{array}{rr}
    1.0 & -0.6  \\
    -0.6 & 1.0   \\
  \end{array}
  \right)
\]


\noindent
Parameters for the 4D study. 
\[
\mu_{\rm MC} = 
\left(
\begin{array}{r}
1.0 \\ 0.0 \\ -0.5 \\ 0.5 \\ 
\end{array}
\right) \ ,
\ \ \
\mu_{\rm true} = 
\left(
\begin{array}{r}
0.8 \\ 0.1 \\ -0.6 \\ 0.7  \\
\end{array}
\right) \ ,
\ \ \
\sigma_{\rm MC} =
\left(
\begin{array}{r}
1.0 \\ 0.7 \\ 1.1 \\ 0.8 \\ 
\end{array}
\right) \ ,
\ \ \
\sigma_{\rm true} =
\left(
\begin{array}{r}
0.8 \\ 0.6 \\ 1.0 \\ 0.6 \\ 
\end{array}
\right) \ ,
\ \ \ 
\sigma_{\rm det} = 
\left(
\begin{array}{r}
0.4 \\ 0.5 \\ 0.6 \\ 0.3 \\ 
\end{array}
\right)
\]
\[
  \rho_{\rm MC} =
  \left(
  \begin{array}{rrrr}
    1.0 & 0.1 & -0.2 & 0.3  \\
    0.1 & 1.0 & 0.0 & 0.1  \\
    -0.2 & 0.0 & 1.0 & 0.7   \\
    0.3 & \ \ 0.1 & 0.7 & \ \ 1.0  \\
  \end{array}
  \right) \ ,
\ \ \ \
  \rho_{\rm true} =
  \left(
  \begin{array}{rrrr}
    1.0 & 0.0 & -0.3 & 0.4  \\
    0.0 & 1.0 & 0.2 & 0.0  \\
    -0.3 & 0.2 & 1.0 & 0.5  \\
    0.4 & \ \ 0.0 & 0.5 & \ \ 1.0  \\ 
  \end{array}
  \right)
\]


\noindent
Parameters for the 6D study. 
\[
\mu_{\rm MC} = 
\left(
\begin{array}{r}
1.0 \\ 0.0 \\ -0.5 \\ 0.5 \\ -1.0 \\ 0.3 \\
\end{array}
\right) \ ,
\ \ \
\mu_{\rm true} = 
\left(
\begin{array}{r}
0.8 \\ 0.1 \\ -0.6 \\ 0.7 \\ -0.8 \\ 0.1 \\
\end{array}
\right) \ ,
\ \ \
\sigma_{\rm MC} =
\left(
\begin{array}{r}
1.0 \\ 0.7 \\ 1.1 \\ 0.8 \\ 1.2 \\ 1.4 \\
\end{array}
\right) \ ,
\ \ \
\sigma_{\rm true} =
\left(
\begin{array}{r}
0.8 \\ 0.6 \\ 1.0 \\ 0.6 \\ 1.0 \\ 1.1 \\
\end{array}
\right) \ ,
\ \ \ 
\sigma_{\rm det} = 
\left(
\begin{array}{r}
0.4 \\ 0.5 \\ 0.6 \\ 0.3 \\ 0.4 \\ 0.4 \\
\end{array}
\right)
\]
\[
  \rho_{\rm MC} =
  \left(
  \begin{array}{rrrrrr}
    1.0 & 0.1 & 0.2 & -0.3 & 0.0 & 0.0 \\
    0.1 & 1.0 & 0.0 & -0.2 & 0.3 & 0.1 \\
    0.2 & 0.0 & 1.0 & 0.1 & -0.2 & 0.3 \\
    -0.3 & -0.2 & 0.1 & 1.0 & 0.1 & 0.0 \\
    0.0 & 0.3 & -0.2 & 0.1 & 1.0 & 0.7 \\
    0.0 & 0.1 & 0.3 & 0.0 & 0.7 & \ 1.0 \\  
  \end{array}
  \right) \ ,
\ \ \ \
  \rho_{\rm true} =
  \left(
  \begin{array}{rrrrrr}
    1.0 & 0.0 & 0.2 & -0.2 & 0.1 & 0.0 \\
    0.0 & 1.0 & 0.0 & -0.1 & 0.2 & 0.0 \\
    0.2 & 0.0 & 1.0 & 0.0 & -0.3 & 0.4 \\
    -0.2 & -0.1 & 0.0 & 1.0 & 0.2 & 0.0 \\
    0.0 & 0.2 & -0.3 & 0.2 & 1.0 & 0.5 \\
    0.0 & 0.0 & 0.4 & 0.0 & 0.5 & \ 1.0 \\  
  \end{array}
  \right)
\]


\section{ Activation function dependence of event correlations }
\label{app:activation}

\begin{figure}[ht!]
   \begin{center}
\includegraphics[width=0.95\linewidth]{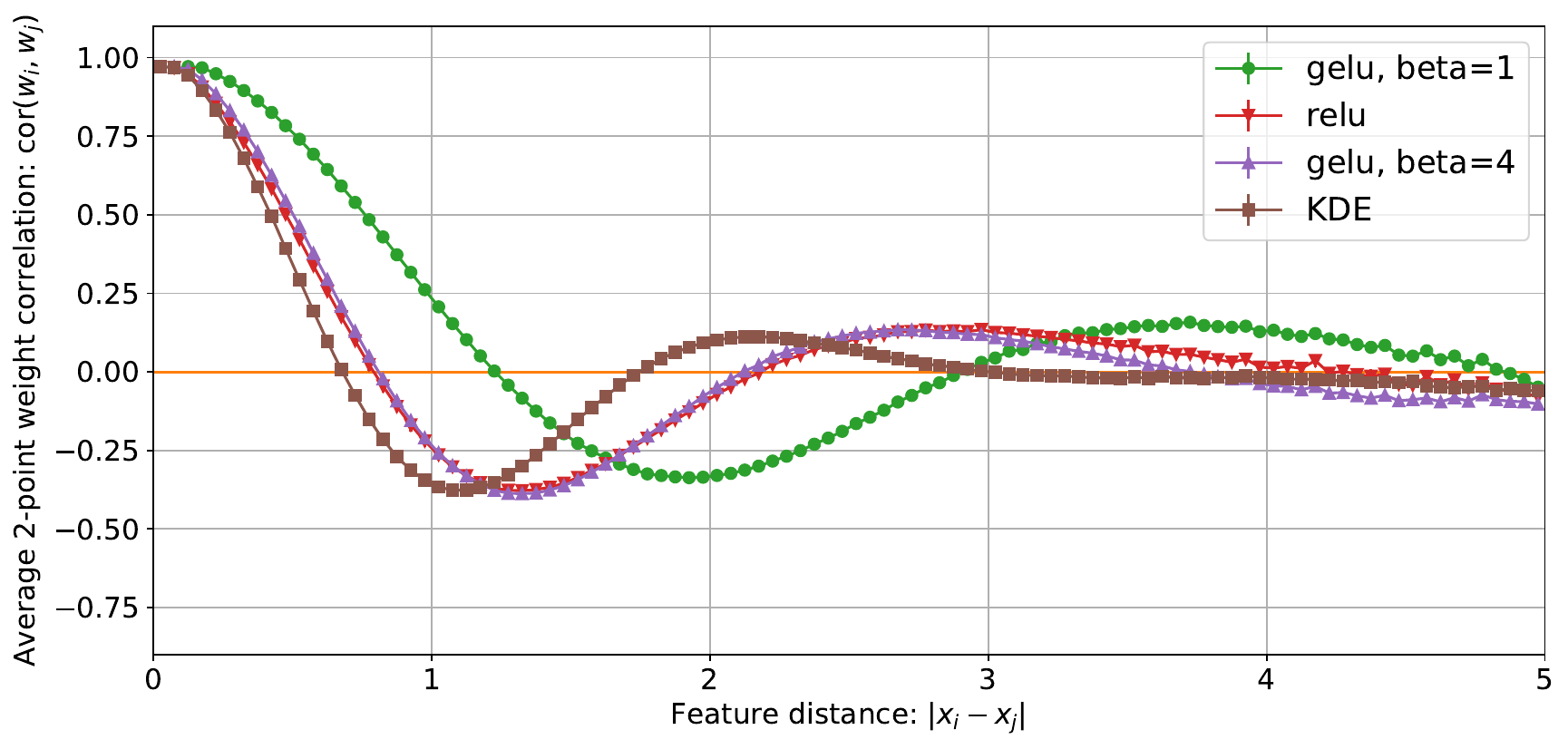}
   \caption{
      Average weight correlation between two events as a function of the absolute distance between the events in the observable.  The results for the KDE are compared with three different NN activation function choices: {\tt relu},
      {\tt gelu} with $\beta=1$, and {\tt gelu} with $\beta=4$.
   }
   \label{fig:weight-correlation-vs-distance-1d-activations}
   \end{center}
\end{figure}

We have found that the event weight correlations in the unfolding output are sensitive to the activation functions used in the NN model.  Figure~\ref{fig:weight-correlation-vs-distance-1d-activations} shows the average event weight correlation as a function of distance in feature space for our 1D study with a detector smearing of 0.5 for the KDE compared to three different choices for the NN activation function: {\tt relu}, {\tt gelu} with $\beta=1$, and {\tt gelu} with $\beta=4$.  We find that {\tt gelu} with $\beta=1$, which is the standard version of {\tt gelu}, introduces the most regularization and has the longest distance scale for the correlations.  If the transition from flat to a slope of 1 is shortened within {\tt gelu} by setting $\beta=4$, the pattern of event weight correlations is quite similar to {\tt relu} and this is similar to the results from the KDE.

\begin{figure}[ht!]
   \begin{center}
\includegraphics[width=0.95\linewidth]{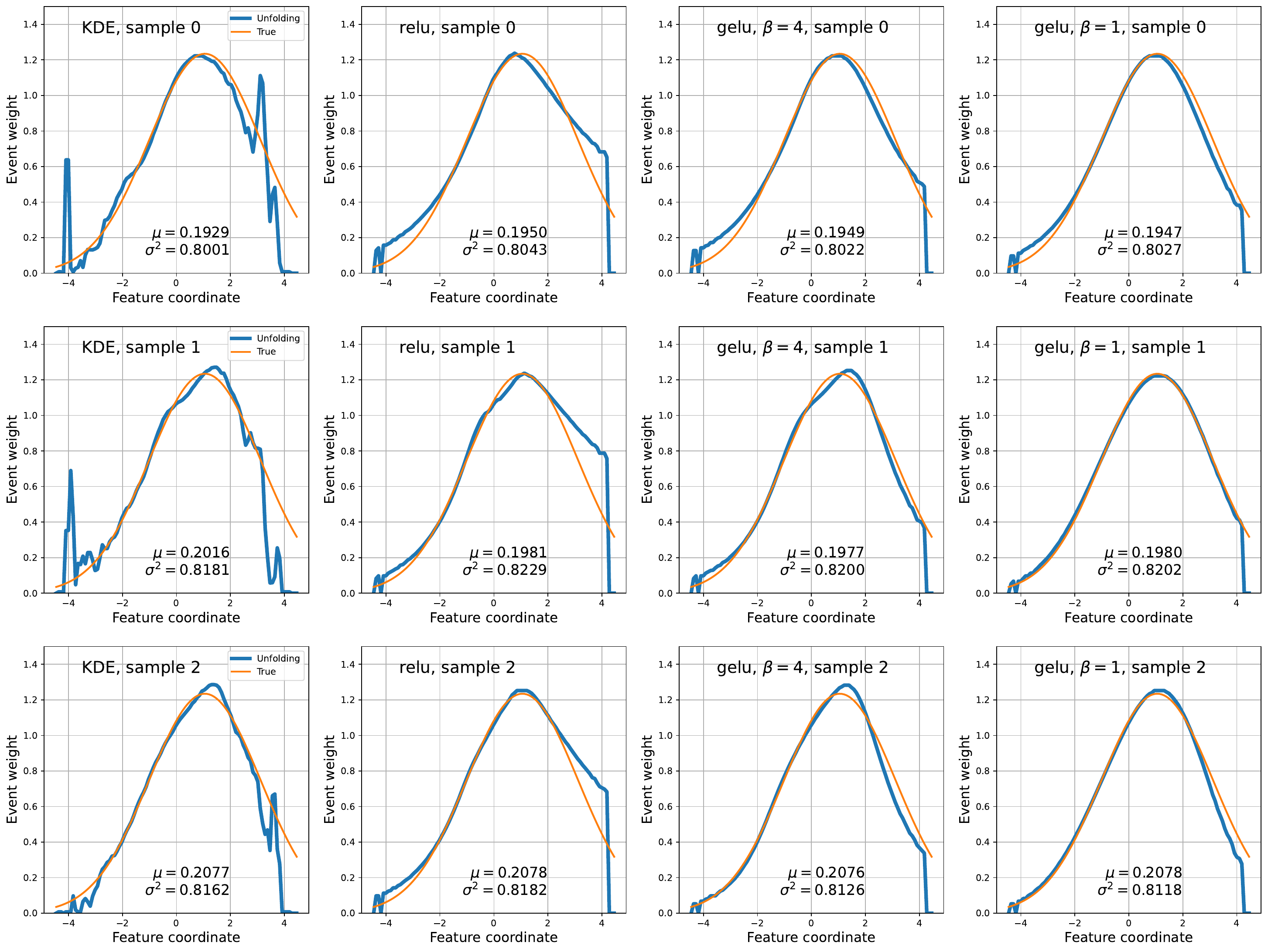}
   \caption{
      Graphs of the event weight function from the unfolding compared with the true event weight function.  The columns show different options within OmniFold: left is the KDE, center left is NNs with {\tt relu} activation, center right is NNs with {\tt gelu} activation with $\beta=4$, and right is NNs with {\tt gelu} activation with $\beta=1$.  Each row is the results for a particular test sample and the columns within a row all use the same test sample.
   }
   \label{fig:weight-functions}
   \end{center}
\end{figure}

Figure~\ref{fig:weight-functions} shows graphs of the event weight function from the unfolding compared with the true event weight function.  The columns compare the KDE with the three NN activation function choices.  Each row shows the results for unfolding a particular test dataset.  The KDE has the least amount of regularization of the weight function.  Statistical fluctuations in the tails are visible.  The {\tt relu} activation allows for sharper kinks in the weight function and tends to converge to a linear function on the right side, where the statistics are lower.  The {\tt gelu} with $\beta=1$ is much more regularized than {\tt gelu} with $\beta=4$, which is closest to a smoothed version of the KDE weight function.

The true weight function is smooth and slowly varying, so a high degree of regularization is not necessarily a problem for our particular toy physics model.  The inference results from the unfolding are given for each plot in Figure~\ref{fig:weight-functions}.  Even though the weight functions differ substantially, they all give very similar values in the inference of the model parameters.  After seeing the graphs for {\tt gelu} with $\beta=1$ (right column in Figure~\ref{fig:weight-functions}), we expected it to have better precision, but that's not the case.  The KDE and all three NN activation choices all give inference uncertainties of around 0.010 for the model $\mu$ and 0.014 for the model $\sigma^2$.  Note that the inference uncertainty scale is quite small compared to the horizontal axis range shown in Figure~\ref{fig:weight-functions}.

\section{ Ensembling and inference resolution from NN model parameter initialization}
\label{app:ensembling}

\begin{figure}[ht!]
   \begin{center}
\includegraphics[width=0.95\linewidth]{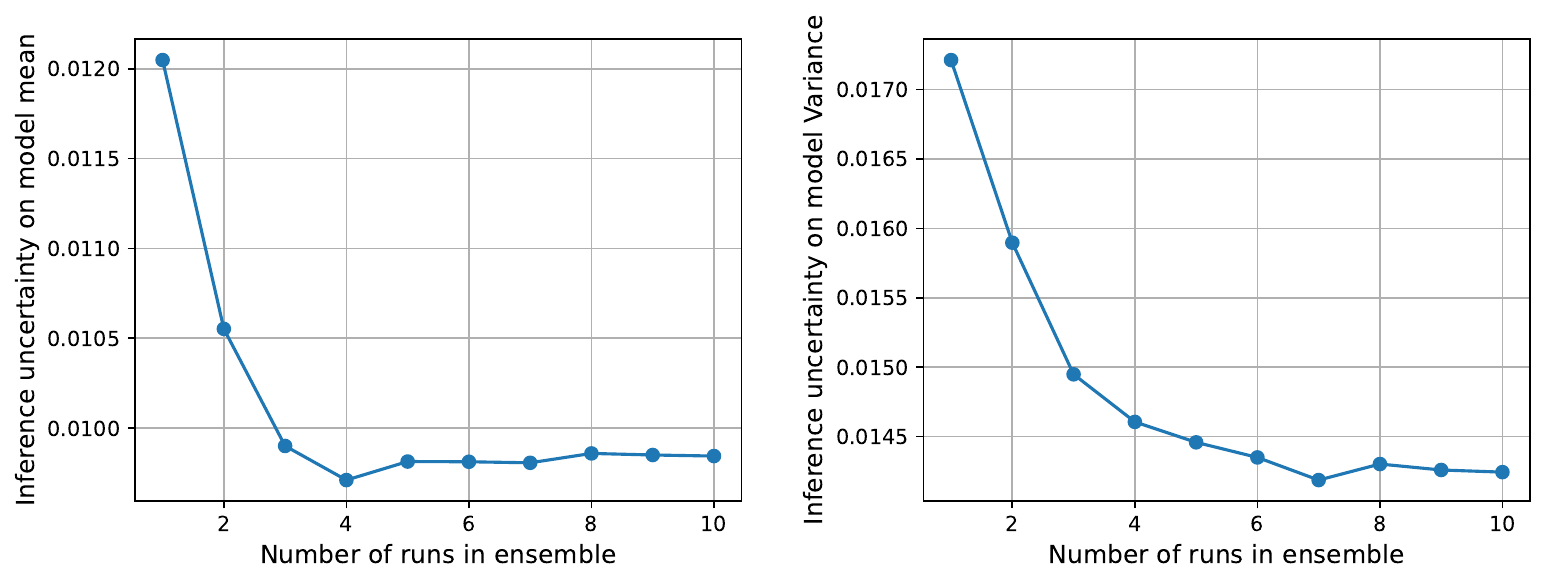}
   \caption{
      Inference uncertainty on the model parameters for the 1D study as a function of the number of runs in the ensemble average.  A detector smearing of 0.5 was used.
   }
   \label{fig:resolution-vs-n-average}
   \end{center}
\end{figure}

The converged state of the unbinned unfolding when using NNs within OmniFold has some dependence on the random initialization of the NN model weights.  This impacts the resolution of the unfolding and downstream inference.  However, this effect can be removed by averaging the output of several runs of the full unfolding, where each run starts with a different set of random initial NN model weights.  This averaging is known as ensembling.  Averaging the final unfolding weights is known as parallel ensembling~\cite{Acosta:2025lsu}.
Figure~\ref{fig:resolution-vs-n-average} shows the inference uncertainty on the model mean and model variance for our 1D study with detector smearing of 0.5 as a function of the number of runs used in the ensemble average.  Our interpretation of these results is that an ensemble of 10 runs is sufficient to remove the parameter initialization contribution to the unfolding resolution.

\begin{figure}[ht!]
   \begin{center}
\includegraphics[width=0.95\linewidth]{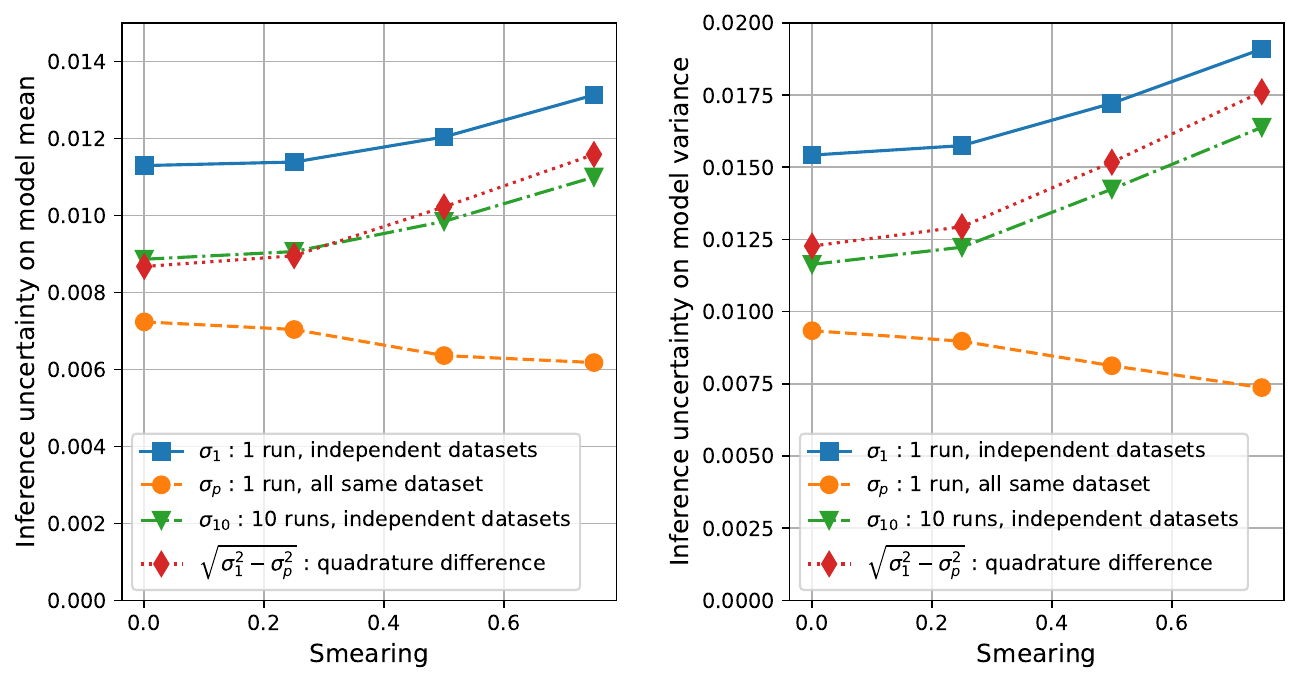}
   \caption{
      Inference uncertainty on the model parameters for our 1D study as a function of detector smearing.  The orange circles are a measure of the contribution from the initialization of the random NN weights $\sigma_p$.  The blue squares show the resolution without ensembling $\sigma_1$.  The Green triangles show the resolution for a 10-run ensemble average $\sigma_{10}$.  The quadrature difference $\sqrt{\sigma_1^2 - \sigma_p^2}$ agrees fairly well with the 10-run ensemble average.
   }
   \label{fig:ensemble-comp-vs-resolution}
   \end{center}
\end{figure}

Figure~\ref{fig:ensemble-comp-vs-resolution} shows the components of the overall inference uncertainty as a function of the level of detector smearing. The orange circles show the resolution, from the RMS of the inference distribution, when the inference is done 200 times on the same dataset with different initial random NN model weights in a single run.  This is a direct measure of the size of the parameter initialization resolution contribution $\sigma_p$.  The blue squares show the results for unfolding 200 independent datasets without ensembling $\sigma_1$.  The green triangles show the results for unfolding the same 200 independent datasets, where the ensemble average of 10 runs is used $\sigma_{10}$.  The red diamonds show the quadrature difference of the 1-run minus the parameter initialization resolution $\sqrt{\sigma_1^2 - \sigma_p^2}$, which agrees fairly well with the 10-run ensemble average resolution $\sigma_{10}$, as expected.

\FloatBarrier
 \bibliographystyle{elsarticle-num} 
 \bibliography{cas-refs,HEPML,other-refs}

\end{document}